\documentclass{iaus}
\usepackage{graphicx}
\def\Rs{R_{\odot}}
\title[Maunder minimum] %% give here short title %%
{Possible explanations of the Maunder minimum from a flux transport dynamo model}
\author[Karak \& Choudhuri]   %% give here short author list %%
{Bidya Binay Karak
 \and Arnab Rai Choudhuri}
\affiliation{Department of Physics, Indian Institute of Science, Bangalore-560012\\ email: {\tt bidya$\_$karak@physics.iisc.ernet.in,} 
\\{\tt arnab@physics.iisc.ernet.in}}

\pubyear{2010}
\volume{xxx}  %% insert here IAU Symposium No.
\pagerange{119--126}
\jname{Title of your IAU Symposium}
\editors{A.C. Editor, B.D. Editor \& C.E. Editor, eds.}
\begin{document}

\maketitle

\begin{abstract}
We propose that at the beginning of the Maunder minimum the 
poloidal field or amplitude of meridional circulation or both fell 
abruptly to low values. With this proposition, a flux transport 
dynamo model is able to reproduce various important aspects 
of the historical records of the Maunder minimum remarkably well.
\end{abstract}

\firstsection % if your document starts with a section,
\section{Introduction}
One important aspect of the solar cycle is the Maunder minimum during 
1645--1715 when the solar activity was strongly reduced 
(Ribes \& Nesme-Ribes 1993). It was not 
an artifact of few observations, but a real phenomenon 
(Hoyt \& Schatten 1996). From the study of historical data 
(Ribes \& Nesme-Ribes 1993), it has been 
confirmed that the sunspot numbers in both the hemisphere 
fell abruptly to nearly zero value at the beginning of the 
Maunder minimum, whereas a few sunspots appeared 
in the southern hemisphere during the last phase. It is also 
established from the cosmogenic isotopes data (Beer et al. 1998; 
Miyahara et al. 2004) that the cyclic oscillations of 
solar activity continued in the heliosphere at a weaker level during the 
Maunder minimum, but with a period of 13--15 
years instead of the regular 11-year period.

The most promising model of studying solar cycle at present is the flux 
transport dynamo model (Choudhuri et al. 1995; Durney 1995;
Dikpati \& Charbonneau 1999; 
Chatterjee et al. 2004). 
The main sources of irregularities in this model are the stochastic 
fluctuations in the Babcock--Leighton process of  poloidal 
field generation (Choudhuri 1992; Choudhuri et al. 2007) and the stochastic 
fluctuations of meridional circulation (hereafter MC) (Hathaway 1996). 
Therefore we propose that the polar field or amplitude of MC or 
both decreased at the beginning of Maunder minimum. With this 
proposition, we use a flux transport dynamo model to reproduce 
a Maunder minimum. The details
of this work can be found in Choudhuri \& Karak (2009) and 
Karak (2010).

\section{Methodology}

We cary out all the analyses with the flux transport dynamo model 
described in Chatterjee et al. (2004). To reproduce the Maunder minimum, 
we perform the following three separate sets of experiments. 
Similar to Choudhuri et al. (2007), first, we decrease the polar field 
above $0.8 \Rs$ by a factor $\gamma$ after stopping the code at 
a solar minimum. We change 
the polar field by different amount in two hemispheres. 
In northern hemisphere, we take $\gamma = 0.0$, whereas 
in southern hemisphere, it is  $0.4$. In addition, 
in this calculation, we decrease the toroidal field by 
multiplying it everywhere by 0.8 to stop the eruption 
for some time. This essentially reduces 
the strong overlap between two cycles in our model (see figure 13 of 
Chatterjee et al. 2004). After making these changes, we run the 
model for several cycles without any further change. 
In the second procedure for reproducing the Maunder minimum, 
we decrease the amplitude of MC $v_0$ abruptly to a very low 
value. After keeping it at low value for few years, we again 
increase it to the usual value but at different rates in two 
hemispheres. In the northern hemisphere, it is increased at slightly 
lower rate than the southern hemisphere. Note that in this case 
we have varied only $v_0$ and no other parameters of the model.
We have repeated this calculation in the low 
diffusivity model of Dikpati \& Charbonneau (1999) too. 
Last, we have included the effect of the fluctuations of 
polar field along with the fluctuations of MC. We have run 
the model for different values of $\gamma$s from 0 to 1 at 
each values of $v_0$ from a very low value to the average 
value. Then we find out the critical values of $v_0$ and 
the corresponding $\gamma$ factor for which we 
get a Maunder-like minimum.

\begin{figure}
\centering{\includegraphics[width=14cm]{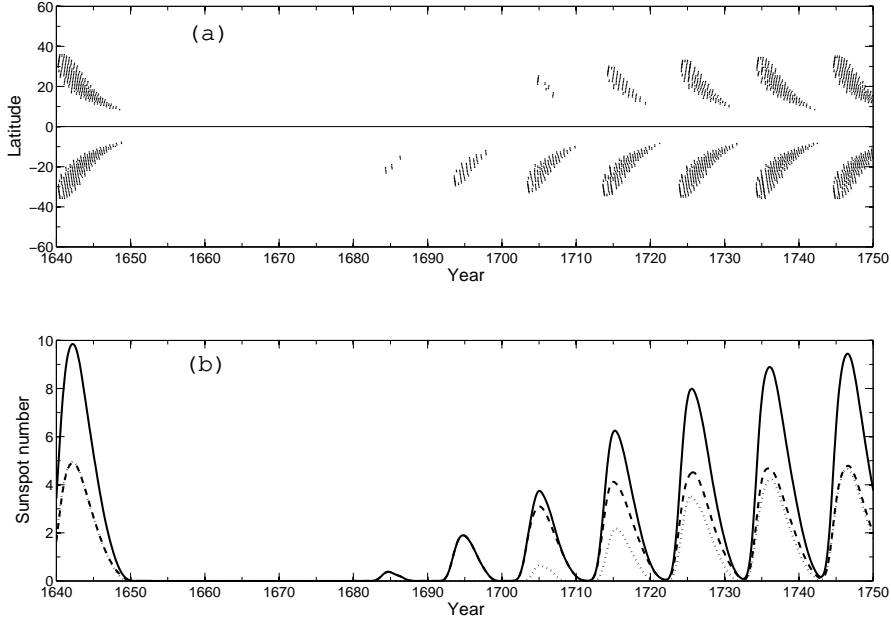}}
\caption{Results covering the Maunder minimum episode. (a) The butterfly 
diagram. (b) The smoothed sunspot number. The dashed and dotted lines show the
sunspot numbers in southern and northern hemispheres, whereas the solid
line is the total sunspot number. From Choudhuri \& Karak (2009).}
\label{mmp}
\end{figure}

\section{Results}

First, we discuss the results from the polar field reduction procedure. It is 
shown in Fig. \ref{mmp} (see the caption also). In order to 
facilitate the comparison with the
observation data, we have marked the beginning of Fig. \ref{mmp}
to be the year 1640.
From this figure 
we see that the sudden initiation but gradual recovery of Maunder 
minimum and the north-south asymmetry of sunspot numbers in the last 
last phase have been nicely reproduced. We also find the cyclic oscillation of 
the poloidal field in the solar wind (shown in figure~2 of Choudhuri \& 
Karak 2009). This oscillation explains the cyclic behavior found in 
cosmogenic isotopes data. In this calculation, we have taken $\alpha$ = 
21 m s$^{-1}$ and turbulent diffusivity $\eta_p$ of the poloidal
field within the convection zone = $3.2 \times 10^{12}$ 
cm$^2$ s$^{-1}$. 
This combination of $\alpha$ and $\eta_p$ 
gives the correct growth rate to produce Maunder minimum. 
In Choudhuri \& Karak (2009), we list some other combinations 
which also reproduce maunder-like grand minima. 

%\begin{figure}
%\centering{\includegraphics[width=16cm]{fig2}}
%\caption{This kind of variation of the amplitude of 
%MC (in m s$^{-1}$) in northern (solid line) and southern 
%hemispheres (dashed line) during Maunder minimum reproduces 
%exactly similar figure as shown in Fig. \ref{mmp}.}
%\label{mmm}
%\end{figure}
\begin{figure}
\centering{\includegraphics[width=8cm]{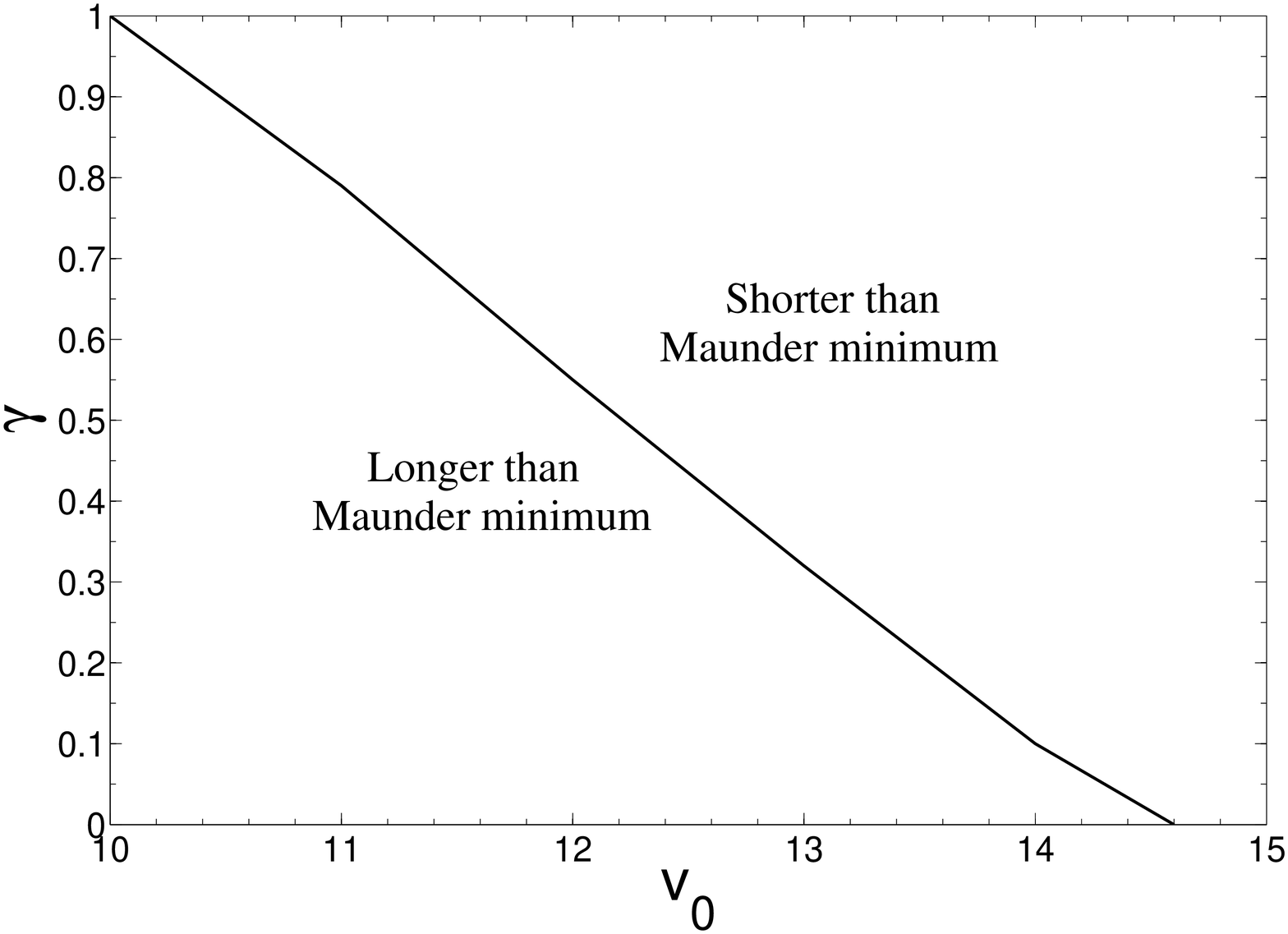}}
\caption{Parameter space of amplitude of MC ($v_0$) and polar field
reduction factor ($\gamma$). The line shows the values of these 
parameters which are giving Maunder-like grand minima.}
\label{para}
\end{figure}

In the second procedure of reducing $v_0$ alone, we again get
results very similar to what are shown in Fig. \ref{mmp}, if
$v_0$ is abruptly made 10 m s$^{-1}$ and then allowed to gain
back its strength (see figures in Karak 2010).
This procedure also reproduces 
all the important features of the Maunder minimum remarkably well. 
We underscore that in the advection-dominated model (e.g. Dikpati \& 
Charbonneau 1999) we do not get this result. This is because 
the decrease of MC in the advection-dominated model produces more 
toroidal field and makes the cycle stronger, as analyzed by 
Yeates et al. (2008).

We conclude that it is possible to reproduce the Maunder minimum 
by decreasing either polar field or $v_0$ to very low values. 
However, if we allow both the polar field and $v_0$ to
decrease simultaneously, then it seems possible to reproduce 
Maunder-like grand minima without making either the polar
field or the MC so low. We have found that for each $\gamma$ there is a 
particular value of $v_0$ which can produce a Maunder-like 
minimum. The line in Fig. \ref{para} shows the combinations 
of parameters giving a Maunder minimum of appropriate duration. 
Values of $\gamma$ and $v_0$ lying in the lower left of
Fig. \ref{para} give grand minima longer than the
Maunder minimum, whereas values lying in the upper right
give shorter minima. 
We should mention that Fig. \ref{mmp} is reproduced 
just by taking $\gamma = 0.2$ (actually $\gamma_N = 0.0$ 
and $\gamma_S = 0.4$ is used) keeping $v_0$ unchanged. 
However, in this calculation, we have to reduce $v_0$ 
to around 13.5 m s$^{-1}$ along with the polar field 
reduction to 0.2 in both hemispheres. 
This is because in earlier calculations 
we had reduced the toroidal field slightly along with the
change of polar field.
Here we have not done this reduction of the toroidal
field. Additionally, here the values of $\alpha$ and $\eta_p$
are slightly different giving different dynamo growth rate.

\section{Conclusion}

We have shown that most of the important features of the  
Maunder minimum can be reproduced quite well by assuming a 
simple ansatz that the polar field or the amplitude of 
MC or both decreased significantly at the beginning of Maunder minimum. 
Because of our lack of knowledge about the physical
conditions at the beginning of the Maunder minimum,
we cannot say how exactly the Sun was driven to the
Maunder minimum. However, we should mention that there are 
several independent studies (Wang \& Sheeley 2003; 
Miyahara et al. 2004; Passos \& Lopeas 2009) 
suggesting that the amplitude of 
MC was weaker during the Maunder minimum. If this happens to
be correct, then this study along with several other studies (Chatterjee et al. 2004;
Chatterjee \& Choudhuri 2006; Jiang et al. 2007; Yeates et al. 2008; 
Goel \& Choudhuri 2009; Karak \& Choudhuri 2011)
indicate that the solar dynamo actually is 
diffusion-dominated and not advection-dominated.

\end{document}